\documentclass[a4paper,aps,prl,reprint,superscriptaddress, twocolumn,preprintnumbers,amsmath,amssymb,nobalancelastpage,10pt, longbibliography]{revtex4-2}
\usepackage{svg}
\usepackage{amsmath}
\usepackage{verbatim}
\usepackage{graphicx}
\usepackage{color}
\usepackage{tikz}
\usepackage{upgreek}
\usepackage{float}

\usepackage[colorlinks=true,citecolor=blue,linkcolor=blue,urlcolor=blue]{hyperref}

\begin{document}
\title{Ultrafast loss of lattice coherence in the light-induced structural phase transition of V$_2$O$_3$}
\author{A. S. Johnson}
\email[]{allan.s.johnson@gmail.com}
\affiliation{ICFO - Institut de Ci\`encies Fot\`oniques, The Barcelona Institute of Science and Technology, Av. Carl Friedrich Gauss 3, 08860 Castelldefels, Barcelona, Spain}

\author{D. Moreno-Menc\'ia}
\affiliation{ICFO - Institut de Ci\`encies Fot\`oniques, The Barcelona Institute of Science and Technology, Av. Carl Friedrich Gauss 3, 08860 Castelldefels, Barcelona, Spain}
\author{E. B. Amuah}
\affiliation{ICFO - Institut de Ci\`encies Fot\`oniques, The Barcelona Institute of Science and Technology, Av. Carl Friedrich Gauss 3, 08860 Castelldefels, Barcelona, Spain}
\affiliation{Department of Physics and Astronomy, Aarhus University, Ny Munkegade 120, 8000 Aarhus C, Denmark.}
\author{M. Menghini}
\affiliation{Department of Physics and Astronomy, KU Leuven, Celestijnenlaan 200D, 3001 Leuven, Belgium}
\affiliation{IMDEA Nanociencia, C/ Faraday 9, 28049, Madrid, Spain}
\author{J.-P. Locquet}
\affiliation{Department of Physics and Astronomy, KU Leuven, Celestijnenlaan 200D, 3001 Leuven, Belgium}
\author{C. Giannetti}
\affiliation{Department of Mathematics and Physics, Universit\`a Cattolica, I-25121 Brescia, Italy}
\affiliation{Interdisciplinary Laboratories for Advanced Materials Physics (I-LAMP), Universit\`a Cattolica, I-25121 Brescia, Italy}
\author{E. Pastor}
\email[]{erpastor@uji.es}
\affiliation{ICFO - Institut de Ci\`encies Fot\`oniques, The Barcelona Institute of Science and Technology, Av. Carl Friedrich Gauss 3, 08860 Castelldefels, Barcelona, Spain}
\affiliation{Institute of Advanced Materials (INAM), Universitat Jaume I, Avenida de Vicent Sos Baynat, s/n 12006, Castell{\'o}, Spain}
\author{S. E. Wall}
\email[]{simon.wall@phys.au.dk}
\affiliation{ICFO - Institut de Ci\`encies Fot\`oniques, The Barcelona Institute of Science and Technology, Av. Carl Friedrich Gauss 3, 08860 Castelldefels, Barcelona, Spain}
\affiliation{Department of Physics and Astronomy, Aarhus University, Ny Munkegade 120, 8000 Aarhus C, Denmark.}

\begin{abstract}
In solids, the response of the lattice to photo-excitation is often described by the inertial evolution on an impulsively modified potential energy surface which leads to coherent motion. However, it remains unknown if vibrational coherence is sustained through a phase transition, during which coupling between modes can be strong and may lead to rapid loss of coherence. Here we use coherent phonon spectroscopy to track lattice coherence in the structural phase transition of V$_2$O$_3$. In both the low and high symmetry phases unique coherent phonon modes are generated at low fluence. However, coherence is lost when driving between the low and high symmetry phases. Our results suggest strongly-damped non-inertial dynamics dominate during phase transition due to disorder and multi-mode coupling.  
\end{abstract}

\maketitle

\begin{figure}
\centering
\includegraphics[width=8.6cm]{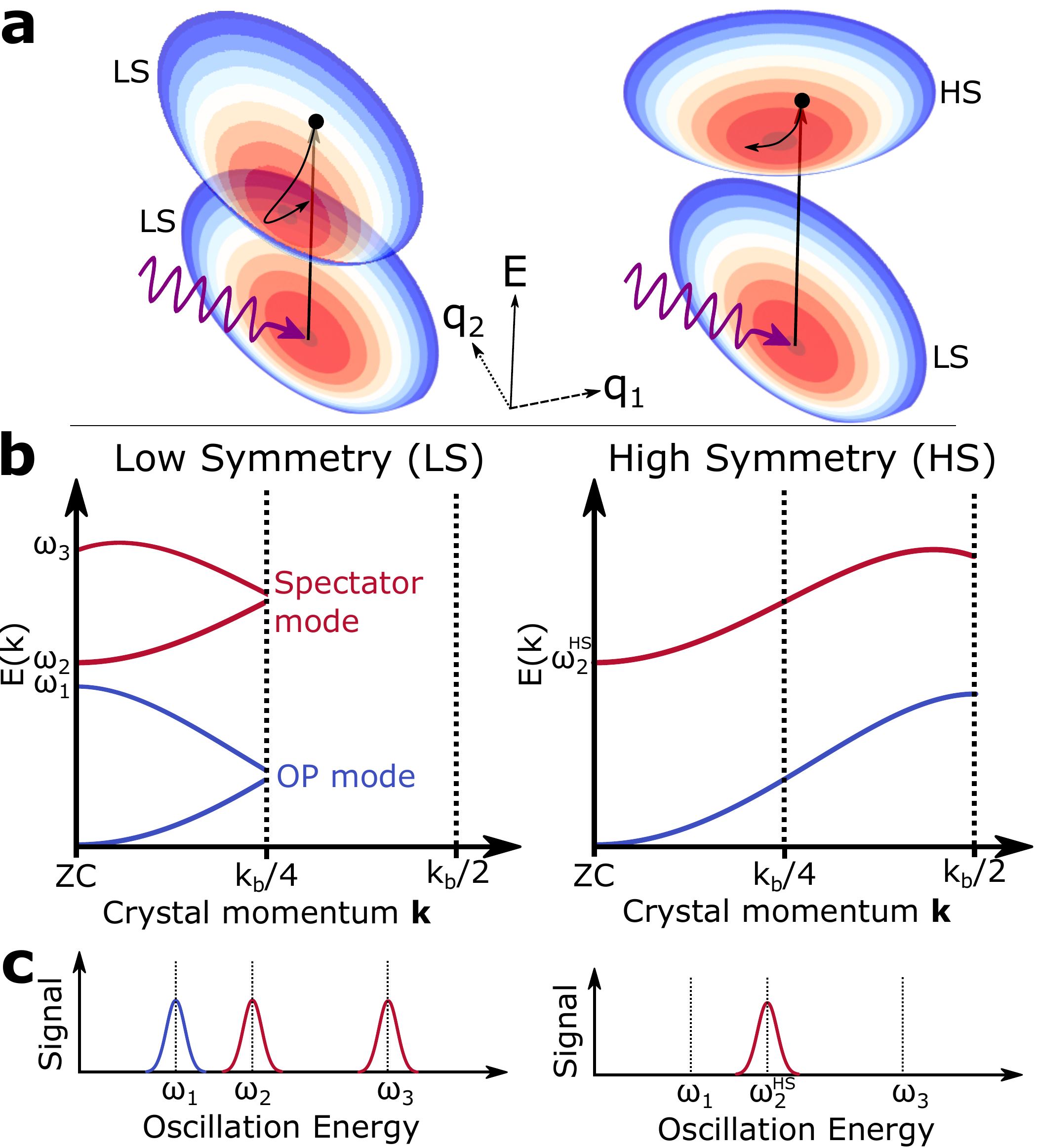}
\caption{(a) Mode potential picture for a phase transition for a lattice potential described by two modes ($q_1, q_2$) following photoexcitation. At low fluence the potential minima shifts promptly, but the symmetry remains, resulting in coherent motion on the LS surface. At high fluence the potential symmetry can also change and in this case the two modes become degenerate. (b) Phonon dispersion picture for a CDW-like transition with an additional optical phonon. The acoustic mode (blue line) drives the CDW transition, creating the amplitude mode at the zone center ($\omega_1$). Due to the change in Brillouin zone associated to the transition from LS to HS, the optical phonon branch (red line) is also  back-folded to create a new mode at the zone center ($\omega_3$). All three zone-centre (ZC) modes can be excited for low fluence excitation in the LS state (c), whereas only $\omega_2^{HS}$ can be excited in the HS state.}
\label{fig1}
\end{figure}

Light-induced phase transitions in quantum materials have the potential to realize novel non-equilibrium phases and transition pathways that differ markedly from their thermal counterparts~\cite{basov_towards_2017,giannetti_ultrafast_2016}. A pathway unique to ultrafast excitation is an impulsive change in the lattice potential, which leads to coherent motion of the lattice on the new potential energy surface (Fig.~\ref{fig1}a). At low fluence, the new potential is a displaced copy of the initial potential and the coherent motion on this surface is defined by the symmetry properties of the initial low symmetry state (LS)~\cite{Zeiger1992}. However, at high fluences, a phase transition can occur to a higher symmetry state (HS). In this case the potential symmetry is changed, possibly to one distinct from the equilibrium HS phase. The resulting coherent response reflects this new symmetry ~\cite{horstmann_coherent_2020}. 

In solids, the change in symmetry often changes the unit cell size and thus the size of the Brillouin zone. This has a direct impact on which modes can be excited (Fig.~\ref{fig1}b). Special attention has been paid to the amplitude mode in case of charge density wave (CDW) systems. In the LS state, low fluence coherently excites the amplitude mode ($\omega_1$ in Fig.~\ref{fig1}b), which lies at the zone centre (ZC). Photo-excitation into the HS state with higher fluence (LS to HS transition) changes the wavevector of the mode to a finite momentum. In this case, the coherent response remains in the system, but the finite momentum modes must be monitored through non-linear effects~\cite{Huber2014, trigo_coherent_2019}.

While the order parameter is the most important mode for the phase transition, quantum materials typically exhibit other phonon modes which change during a phase transition. For example, optical phonons in the HS state (e.g $\omega_2^{HS}$ in Fig.~\ref{fig1}b) back-fold to generate new modes at the zone center in the LS state ($\omega_3$ in Fig.~\ref{fig1}b)~\cite{tomeljak_dynamics_2009}. These modes can be considered spectator modes of the transition~\cite{wegkamp_ultrafast_2015}, as they are affected by the symmetry change, but do not drive the phase transition. Like the amplitude mode, the back-folded optical phonons can be excited coherently in the LS state but, importantly, they have a different response when the phase transition is driven. In VO$_2$, for example, the zone center modes are excited coherently for low fluences in the LS phase~\cite{wall_ultrafast_2012, Wall2013}, but when the HS state is induced the coherence is lost, and the mode transferred to the zone boundary is not observed ~\cite{Wall2018a}.

However, it is unclear how the transition will affect modes that lie at the zone centre of both the LS and HS phases. The question we address here is: \emph{can Raman active modes in the HS state be directly excited from a LS potential during a structural phase transition, as expected from a prompt change in lattice potential?} Answering this question will allow us to determine if disorder is localized in modes that change their wavevector during the phase transformation, or if the disorder spread through all modes of the system. To address this question we use coherent phonon spectroscopy and examine the photo-induced transition in V$_2$O$_3$. Coherent Raman actives modes can be observed optically in both the low and high temperature phases~\cite{Misochko1998, moreno-mencia_ultrafast_2019}, unlike in VO$_2$~\cite{Wall2013}. At low temperature V$_2$O$_3$ is in a monoclinic insulating phase (space group $C2/c$) with 15 Raman active phonons ($7 A_g, 8 B_g$), while above T$_\mathrm{c}=145$\,K~\cite{ronchi_early-stage_2019} the structure changes to corundum (space group $R\bar{3}c$), with 7 Raman active modes ($2 A_{1g}, 5 E_g$) and the system becomes metallic. We focus on the Raman spectra in the 5-12 THz region. Static Raman spectroscopy measurements have shown that the LS spectra consists of three modes at $\omega_1 = 7$\,THz, $\omega_2 =8.29$\,THz and $\omega_3 =9.83$\,THz, which are independent of temperature up to T$_c$~\cite{kuroda_raman_1977, Misochko1998, Tian2016}. Above T$_c$, in the HS state, these three modes are replaced by a single mode at $\omega_2^{HS} =7.5$\,THz which shows broadening and red-shift as the temperature is increased. We aim to coherently excite the $\omega_2^{HS}$ mode when starting in the LS structure.

\begin{figure}
\centering
\includegraphics[width=8.6cm]{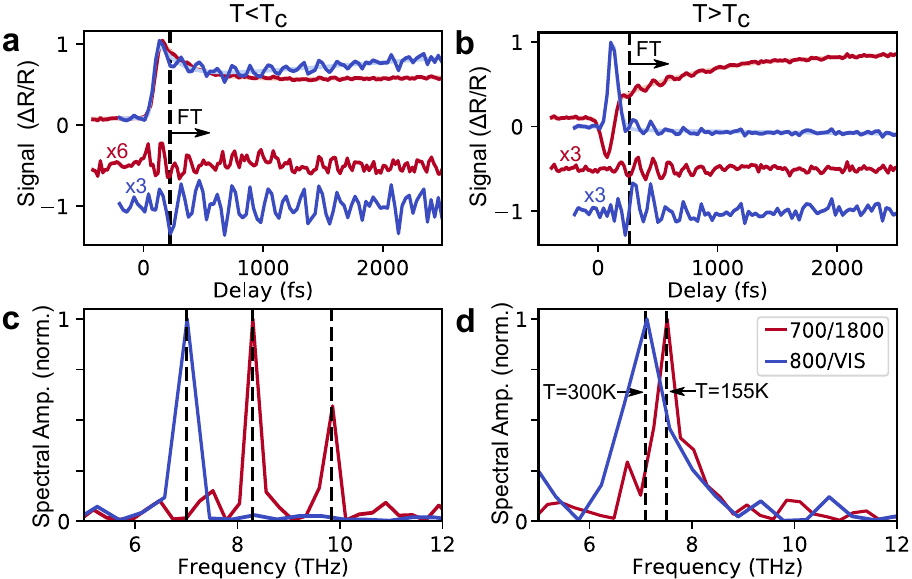}
\caption{Ultrafast measurement of coherent lattice motion in V$_2$O$_3$. 800\,nm pump, broadband visible probe data is shown in blue while 700\,nm pump, 1800\,nm probe data is shown in red. (a) Low fluence excitation pump-probe traces below T$_\mathrm{c}$ (77\,K), along with fits and the fit-subtracted oscillations. Coherent phonon oscillations are extracted taking the Fourier transform from the vertical dashed line onward. (b) Low fluence excitation, above T$_\mathrm{c}$ (300\,K for blue, 155\,K for red). (c) Corresponding spectral amplitude from the Fourier transform of part a. Dashed lines correspond to literature Raman values of the $A_g$ modes. (d) Spectral amplitude above T$_\mathrm{c}$. Indicated are the Raman $A_{1g}$ mode values at 155\,K and 300\,K. }
\label{fig2}
\end{figure}

We study a 56\,nm-thick single-crystalline thin film of V$_2$O$_3$ deposited on a sapphire substrate with a 60\,nm Cr$_2$O$_3$ buffer layer previously described in reference \cite{ronchi_light-assisted_2021} and grown according to the method in reference \cite{dillemans_evidence_2014}. We characterise the ultrafast response in two-different configurations (see Fig.~\ref{fig2}), either pumping with short (15\,fs), 700\,nm laser pulses and probing in the IR with a short (15\,fs) 1800\,nm pulse at 1 kHz repetition rate (red traces), or pumping with a 30\,fs, 800\,nm pulse and probing in the visible with frequency-resolved white-light supercontinuum (Fourier transform limit 15\,fs, 500-750\,nm) at 5 kHz repetition rate (blue traces). The use of multiple probes allows us to achieve greater sensitivity to the phonon modes, as each mode may modulate a different spectral region~\cite{pastor_nonthermal_2022, ramos-alvarez_probing_2019, novelli_localized_2017}. 

We start by examining the low fluence regime, where we measure dynamics on the equilibrium potential. Fig.~\ref{fig2}a shows the reflectivity change at 77\,K, below  T$_\mathrm{c}$, when the sample is in the insulating state after photoexcitation with 2\,mJ/cm$^2$ at 700\,nm, or 6\,mJ/cm$^2$ at 800\,nm. Following an initial fast signal change, the transient response is modulated by oscillations due to coherent phonons. The Fourier transform of the oscillating component is shown in Fig.~\ref{fig2}c. The IR probe shows phonon modes $\omega_2$ and $\omega_3$ at 8.3\,THz and 9.8\,THz, while the visible probe displays $\omega_1$ at 7.0\,THz, in good agreement with the literature values obtained by Raman spectroscopy for the three lowest frequency $A_g$ modes \cite{kuroda_raman_1977}. In agreement with these previous Raman measurements, we observe no shift in the oscillation frequency if the temperature is changed but remains below  T$_\mathrm{c}$ (Fig.~S1). Similarly, Fig.~\ref{fig2}b shows the transient response when the system is in the metallic state at 155\,K (700/1800\,nm, 2.5\,mJ/cm$^2$) or 300\,K (800\,nm/VIS, 10\,mJ/cm$^2$). Here the lowest $A_{1g}$ mode $\omega_2^{HS}$ is observed (Fig.~\ref{fig2}d), whose frequency shifts from 7.5\,THz at 155\,K to 7.1\,THz at 300\,K as expected from the temperature dependence of the Raman mode in the HS phase~\cite{kuroda_raman_1977}.  
\begin{figure*}
\centering
\includegraphics[width=17.2cm]{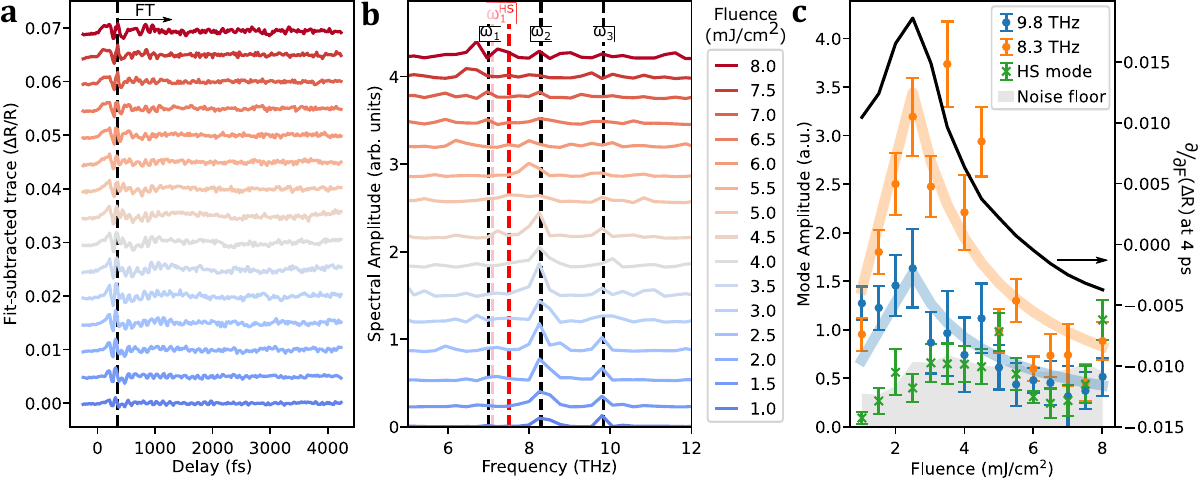}
\caption{Ultrafast lattice decoherence across the phase transition in V$_2$O$_3$. (a) Fit-subtracted coherent phonon oscillations as a function of driving fluence with 700\,nm pump, 1800\,nm probe at 77\,K (traces offset for clarity). The oscillations damp rapidly with increasing fluence and are completely lost at higher fluences. (b) Fourier transform of the data in part a (obtained from the dotted line in (a) onward). Marked as dashed lines the LS phonon frequencies (black, $\omega_1, \omega_2, \omega_3$) and the HS frequencies at 155\,K (red, $\omega_1^{HS}$) and 300\,K (pink). (c) Amplitude of the key phonon modes as a function of fluence. Error bars represent the standard deviation over 25 acquisitions. The grey shaded region denotes the noise level, taken as the maximum of the Fourier transform in the high frequency region 15-25\,THz, in which no signal is expected as it is beyond our temporal resolution. The HS mode frequency is allowed to vary freely between 7.1 and 7.5 THz. Shaded lines indicate fits to our phase mixed model. The black line denotes derivative with respect to fluence of the change in reflectivity observed at 4 ps delay. Both fits and reflectivity indicate a phase transition at around 2.5 mJ/cm$^2$ fluence, above which coherent phonons are no longer excited.}
\label{fig3}
\end{figure*}

Next, we evaluate if the high symmetry mode can be induced by driving the low symmetry initial state. As shown in Fig.~\ref{fig1}, at high fluences, we expect a prompt change in the potential symmetry of the lattice, resulting in a suppression of the low symmetry $Ag$ modes, and the appearance of the high symmetry $A_{1g}$ mode. To test this scenario, we start at 77\,K, and excite with progressively increasing fluence. It is important to begin at temperatures well below T$_\mathrm{c}$ to avoid co-existence of the HS phase in the initial state. As mode $\omega_2$ is the key mode to track, we focus on the IR response.

Fig.~\ref{fig3}a shows the oscillating component of the transient reflectivity at different excitation fluences (raw data in Fig.~S2). At low fluences we observe strong oscillations, which first increase in amplitude as the fluence is increased, before sharply decreasing in lifetime. The signal then saturates, with a few cycles of phonon oscillation visible around 800 fs delay even at the highest fluences. Fig.~\ref{fig3}b shows the corresponding Fourier transform, demonstrating how the coherent phonon peaks at 8.3 and 9.8\,THz are lost to a featureless noise-background as the fluence is increased, without a corresponding appearance of the $\omega_2^{HS}$ between 7.1 and 7.5 THz. We can rule out damage as the low-fluence phonon spectrum remains unchanged even after exposure to  fluences of 16\,mJ/cm$^2$ (Fig.~S3). No shift in frequency is observed, as expected from thermal Raman spectroscopy \cite{kuroda_raman_1977, Tian2016}. 

The amplitude of the phonon modes is plotted in Fig.~\ref{fig3}c, which shows that while the $\omega_2^{HS}$ is not excited at any fluence, the LS modes show an inflection of their behaviour at  2.5\,mJ/cm$^2$, after which their amplitude begins to decrease markedly. A similar inflection is observed in the the derivative of the reflectivity with respect to fluence, as shown in Fig.~\ref{fig3}c. Such a change in the fluence dependence of reflectivity is often taken as the sign of an optically induced phase transition~\cite{vidas_does_2020,ocallahan_inhomogeneity_2015}. To understand the gradual decrease in the phonon amplitude above this fluence and the corresponding reflectivity behaviour we consider a simple model for a first order phase transition taking into account the finite penetration depth of the pump and probe. In this model, below a critical fluence threshold, $F_c$, the LS phonon modes increase in amplitude linearly as the fluence is increased, but at $F_c$ these modes abruptly switch to zero amplitude and instead the HS modes begin to linearly increase. Due to the finite penetration depth of the pump and probe, different regions of each phase can co-exist in the sample and contribute to the total signal; further details in SI section 1. We fit this simple model to our data in Fig.~\ref{fig3}c. The agreement between the reflectivity, phonons, and our simple model is striking, and identifies a fluence threshold of $\sim$2.5\,mJ/cm$^2$ for the phase transition beyond which coherent phonons are not longer excited. The persistent oscillations seen in Fig.~\ref{fig3}a at the highest fluences thus result simply from probing the un-transformed depths of the sample. We reiterate again that, critically, we do not observe the HS mode at any fluence above the fluence threshold. Similar results are obtained shifting the probe frequency to 1600\,nm (Fig.~S4), showing that coherence is not transferred across the light-induced phase transition, even for modes at the zone center. 

\begin{figure}
\centering
\includegraphics[width=8.6cm]{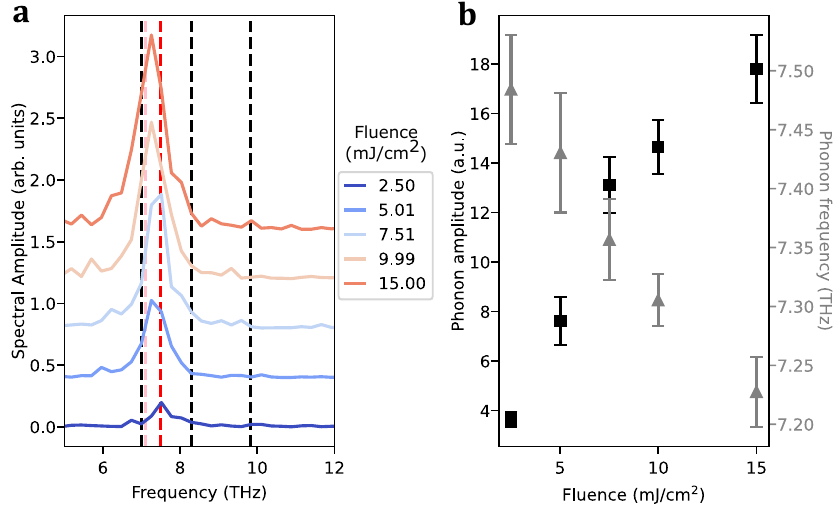}
\caption{Ultrafast lattice dynamics in the high temperature phase of V$_2$O$_3$. (a) Fourier transform of the fit-subtracted transient reflectivity as function of driving fluence (700\,nm pump, 1800\,nm probe) at 155\,K. Traces offset for clarity, raw data shown in the SI. Black dashed lines indicate the LS phonons while the HS mode frequency is highlighted in red for $T=155$\,K and pink for $T=300$\,K. (b) Amplitude and frequency of the phonon mode as a function of fluence. Error bars represent the standard deviation over 25 acquisitions.}
\label{fig4}
\end{figure}

This behaviour is markedly different from the fluence dependence of the HS phase, as shown in Fig.~\ref{fig4} (raw data Fig.~S5), where increasing the fluence up to 15\,mJ/cm$^2$ results in a monotonically increasing amplitude of the oscillation. We also observe a slight red-shift in the frequency, consistent with an increase in the temperature. 

The lack of coherent response in the highly excited LS phase is in contrast to the simple picture presented in Fig.~\ref{fig1} and measurements in CDW transitions~\cite{Huber2014, trigo_coherent_2019}. We now consider different scenarios to explain the lack of coherence transfer during the phase transition. One possibility could be that the lattice potential change may not be sufficiently fast to trigger coherent motion. However, this seems unlikely as the ultrafast loss of vibrational coherence in VO$_2$ coincided with a ultrafast change in the lattice symmetry~\cite{Wall2018a}. 

An alternative scenario is one in which the lattice symmetry change is prompt, but dephasing dominates and the dynamics become overdamped. While an increased coupling to the photoexcited charges, resulting in more scattering, could be the origin for the increased dephasing, it seems unlikely as the HS mode can be clearly observed when strongly photoexciting the metallic state, where the number of free electrons should be even higher. Therefore, it seems most likely that phonon-phonon coupling drives the dephasing and energy is rapidly transferred to modes throughout the Brillouin zone. In addition, as the LS state is anti-ferromagnetic, spin-phonon coupling may also play a role~\cite{Wall2009, Hu2019}. This suggests that the light-induced transition is of the order-disorder type. Instead of atoms moving in a coordinated manner along a displacive coordinate with a fixed wavevector, they are scattered randomly, by phonons, electrons or spins, resulting in uncorrelated motion that occur without a well-defined wavevector.

It is worth contrasting V$_2$O$_3$ with other materials that have been shown to undergo ultrafast light-induced disorder transitions, VO$_2$~\cite{wall_ultrafast_2012, Wall2013, Wall2018a} and La$_{0.5}$Sr$_{1.5}$MnO$_4$ (LSMO) \cite{perez-salinas_multi-mode_2022}. All three have different crystal and spin structures, but the HS phases of both VO$_2$ and LSMO have highly damped phonon modes that prevent clear coherent signals being generated even when beginning in the HS state. Thus it may not be surprising that coherence is not maintained across these phase transitions. However, this is not the case for V$_2$O$_3$, where the HS phonon is robust and damping in the HS state is not an issue. Instead, our work points to the importance of coupling between the modes of the high and low symmetry states during the transition. While the backfolding sketched in Fig.~\ref{fig1} explains why the number of modes changes at the zone center during the phase transition, it does not take into account coupling between the modes. Phonon-phonon interactions mean that the LS modes can be a complex super-position of many HS modes, and this mixing could drive the rapid loss of coherence, even if the HS potential is not strongly damped. 

Furthermore V$_2$O$_3$ undergoes a large discontinuous change in volume across the phase transition~\cite{mcwhan_metal-insulator_1970}, as is typical for many phase-change materials. Initially, the unit cell symmetry changes on the ultrafast timescale, but the unit cell volume cannot react as fast~\cite{singer_nonequilibrium_2018}, and thus, after photoexcitation, the metallic phase is effectively compressed. This could lead to increased lattice anharmonicity and multi-mode scattering out-of-equilibrium when compared to the thermal transition. The increased strength of the mode coupling in the photoexcited state will then persist until the unit cell fully relaxes after several hundreds of picoseconds. Such a process could be tracked in a three pulse measurement~\cite{Wall2013}. As volume changes are universal across structural phase transitions, our results may suggest that disorder-type transitions may be the norm in light-induced phase transitions. This will be true even in systems without strong phonon anharmonicity in equilibrium. Moreover our results confirm that disorder is not only limited to modes that change their wavevector through the transition but can affect all modes in the system.

Finally, we note that we cannot conclusively prove that the photoexcited HS state is the same as the high temperature thermal phase without a direct probe of the lattice. Regardless, the lattice dynamics are likely dominated by disorder and overdamping. In principle, a high symmetry state could be generated in which there are no Raman active modes and in this case the lattice response could remain harmonic. However, in this case the transient symmetry would have to be significantly higher than both the equilibrium low and high temperature structures. Such a high symmetry phase should then also be accessible from the high temperature structure, which is not seen. Ultimately, time-resolve diffuse X-ray scattering would be needed to definitively prove the symmetry and disorder in the photoinduced state~\cite{Wall2018a}. 

In conclusion, we find that lattice coherence is lost when exciting the insulator to metal phase transition of V$_2$O$_3$. The lack of coherence in the transition suggests that strong mode coupling results in overdamped dynamics with energy rapidly transferred to multiple degrees of freedom. Our data suggests the light-induced phase transition in V$_2$O$_3$ is disorder-driven, as in VO$_2$~\cite{Wall2018a} and La$_{0.5}$Sr$_{1.5}$MnO$_{4}$~\cite{perez-salinas_multi-mode_2022}, and further highlights that materials which undergo first-order transitions may be dominated by disordering processes during ultrafast photo-induced phase transitions. This would place major limitations on our ability to coherently control phase transitions. Further work is needed to understand how disorder emerges and whether it can be controlled~\cite{picano_inhomogeneous_2021}.

\begin{acknowledgments}
This work was funded through the European Research Council (ERC) under the European Union’s Horizon 2020 Research and Innovation Programme (Grant Agreement No. 758461) and PGC2018-097027-B-I00 project funded by MCIN/ AEI /10.13039/501100011033/ FEDER `A way to make Europe' and CEX2019-000910-S [MCIN/ AEI/10.13039/501100011033], Fundaci\'o Cellex, Fundaci\'o Mir-Puig, and Generalitat de Catalunya through the European Social Fund FEDER and CERCA program (AGAUR Grant No. 2017 SGR 134, QuantumCAT \ U16-011424, co-funded by ERDF Operational Program of Catalonia 2014-2020). E.P acknowledges the support form IJC2018-037384-I funded by MCIN/AEI /10.13039/501100011033. ASJ and EP acknowledge support of fellowships from `la Caixa' Foundation (ID 100010434), fellowship codes LCF/BQ/PR21/11840013 and LCF/BQ/PR22/11920013. EP and ASJ also thank support from the Marie Sk\l{}odowska-Curie grant agreement no. 754510 (PROBIST). ASJ also acknowledges support from: ERC AdG NOQIA; Agencia Estatal de Investigación (Plan National FIDEUA PID2019-106901GB-I00, FPI, QUANTERA MAQS PCI2019-111828-2, Proyectos de I+D+I “Retos Colaboración” QUSPIN RTC2019-007196-7) ; EU Horizon 2020 FET-OPEN OPTOlogic (Grant No 899794); National Science Centre, Poland (Symfonia Grant No. 2016/20/W/ST4/00314); European Union’s Horizon 2020 research and innovation programme under the Marie-Skłodowska-Curie grant agreement No 101029393 (STREDCH) and No 847648 (“La Caixa” Junior Leaders fellowships ID100010434: LCF/BQ/PI19/11690013, LCF/BQ/PI20/11760031, LCF/BQ/PR20/11770012, LCF/BQ/PR21/11840013).

\end{acknowledgments}
\bibliography{V2O3refs.bib}
\clearpage
\renewcommand{\thefigure}{S\arabic{figure}}
\setcounter{figure}{0}    

\vspace{2cm}
\onecolumngrid
\begin{center}
\textbf{\large Supplemental Material}
\vspace{0.5cm}
\end{center}
\twocolumngrid
\textbf{S1: Penetration depth effects}\\
We model the penetration depth effects of the phase transition as follows. The pump excites the sample with an exponentially decaying profile $F(x)$ with penetration depth $\alpha_{pump}$. The probe pulse then samples both the initial metallic layer where $F(x)$ is greater than the critical fluence $F_c$, and the remaining insulating fraction. We calculate the field amplitudes in each region as follows, enforcing continuity at the insulator-metal boundary but ignoring internal reflection:
\begin{align}
    F(x)&=F_0 e^{-x\alpha_{pump}}\\
    Pr_{met}&=e^{-x\alpha_{met}}(H(F(x)-F_c))\\
    Pr_{ins}(x)&=e^{-x\alpha_{ins}}(1-H(F(x)-F_c))\\
    &\times e^{-x\alpha_{met}}\delta(F(x)-F_c). \notag
\end{align}
From this we can calculate the measured mode amplitudes for the phonons:
\begin{align}
    A_{\omega_i}(x)&=a_{\omega_i}\times F(x) H(F(x)-F_c)\\
    A_{\omega_i,tot}&=\int A_{\omega_i}(x)Pr(x) dx
\end{align}
where $a_{\omega_i}$ is the mode amplitude coefficient and $Pr(x)$ is the probe amplitude  from the corresponding phase. We fit the two LS mode amplitudes ($a_{\omega_i}$s) and the probe penetration depths for the insulating ($\alpha_{ins}$) and metallic ($\alpha_{met}$) regions as a function of input fluence $F_0$ to the data in Fig.~3c. Additional fitting of the noise limited HS mode was not found to significantly impact the results.

\begin{figure}
\centering
\includegraphics[width=8.6cm]{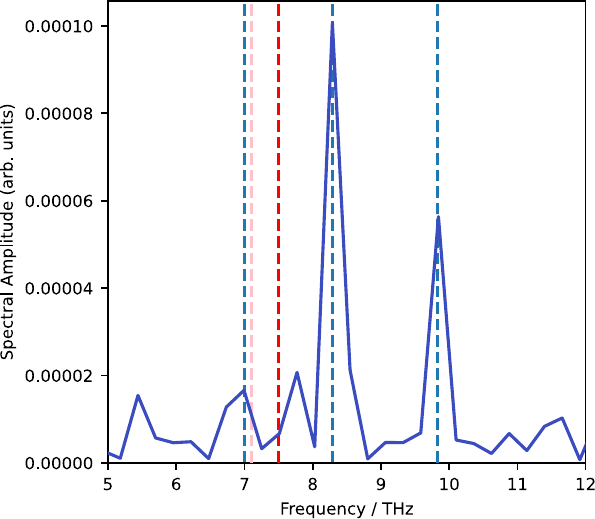}
\caption{Coherent phonon spectra at 25\,K in V$_2$O$_3$. Data taken in the 700\,nm pump, 1800\,nm probe configuration with a fluence of 2.5 mJ/cm$^2$. The phonon modes show no shift from their values at 77K.}
\label{SI_25K}
\end{figure}

\begin{figure}
\centering
\includegraphics[width=8.6cm]{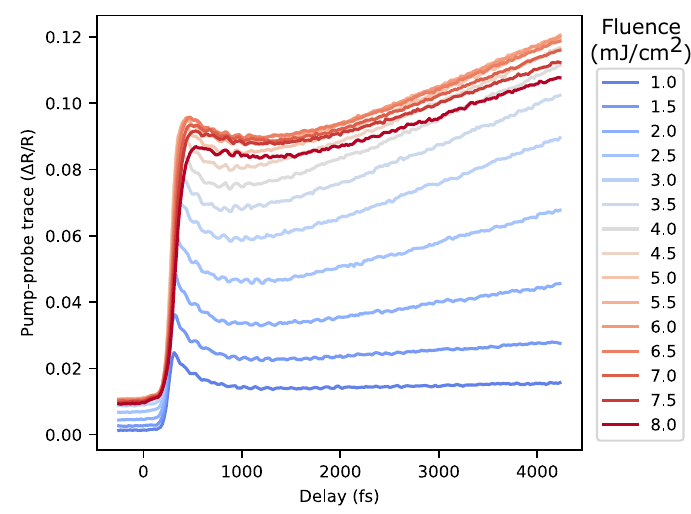}
\caption{Full pump-probe trace as a function of fluence at 77\,K in V$_2$O$_3$ for the data shown in Fig.~3. Data taken in the 700\,nm pump, 1800\,nm probe configuration. Also shown are the fits (faded lines below data), which are barely perceivable due to the quality of the fits. The fit function used for all time-dependent data in the manuscript is $\Delta R/R = \frac{1}{2}(1+\mathrm{erf}(t/\tau_p))(Ae^{-t/\tau_A}+Be^{-t/\tau_B}+C)$.}
\label{SI_77K}
\end{figure}

\begin{figure}
\centering
\includegraphics[width=8.6cm]{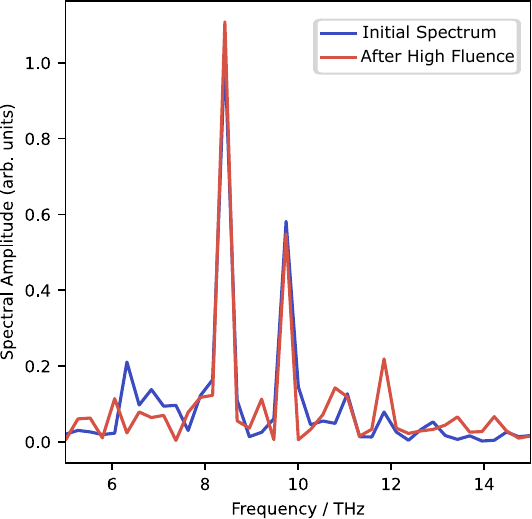}
\caption{Coherent phonon spectrum in V2O3 at 77K before and after prolonged exposure to 16 mJ/cm$^2$ excitation. Data taken in the 700\,nm pump, 1800\,nm probe configuration with a fluence of 1 mJ/cm$^2$. }
\label{SI_repeatability}
\end{figure}

\begin{figure*}
\centering
\includegraphics[width=17.2cm]{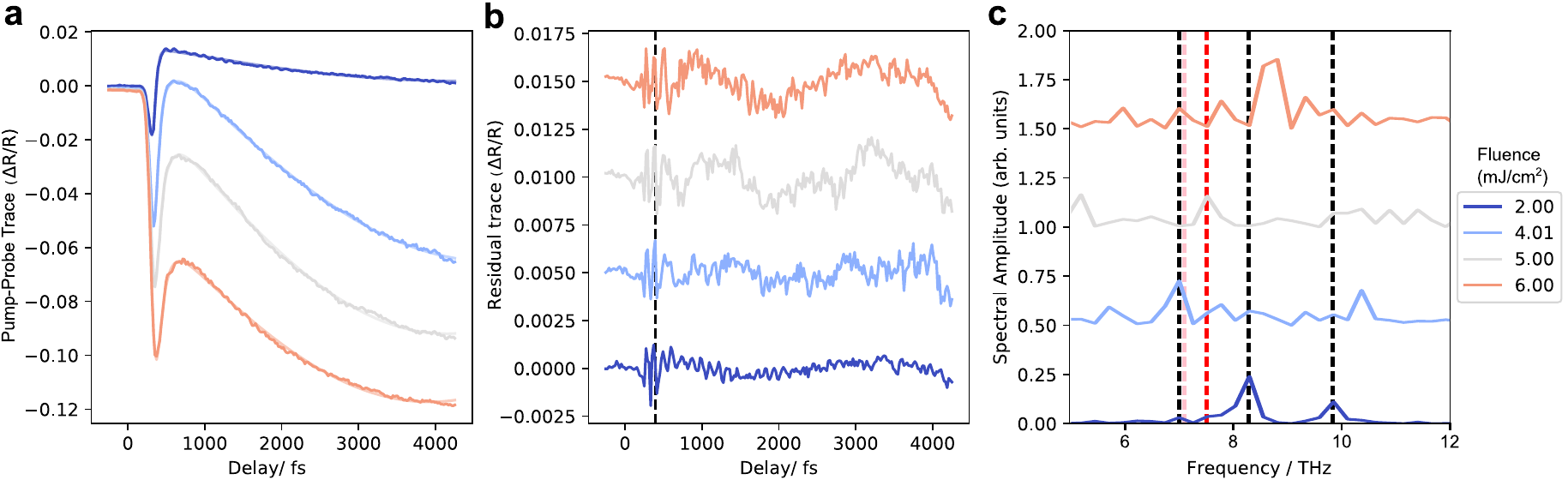}
\caption{1600\,nm probe data at 77\,K. Data taken in the 700\,nm pump, 1600\,nm probe configuration. (a) Full pump probe traces as a function of fluence, along with fits (faded lines below data). (b) Fit-subtracted traces showing clear oscillations at low fluence. (c) Fourier transform of the fit-subtracted traces. The two characteristics phonons are clearly visible at the lowest fluence but are lost as the fluence is increased.}
\label{SI_1600nm}
\end{figure*}

\begin{figure*}
\centering
\includegraphics[width=17.2cm]{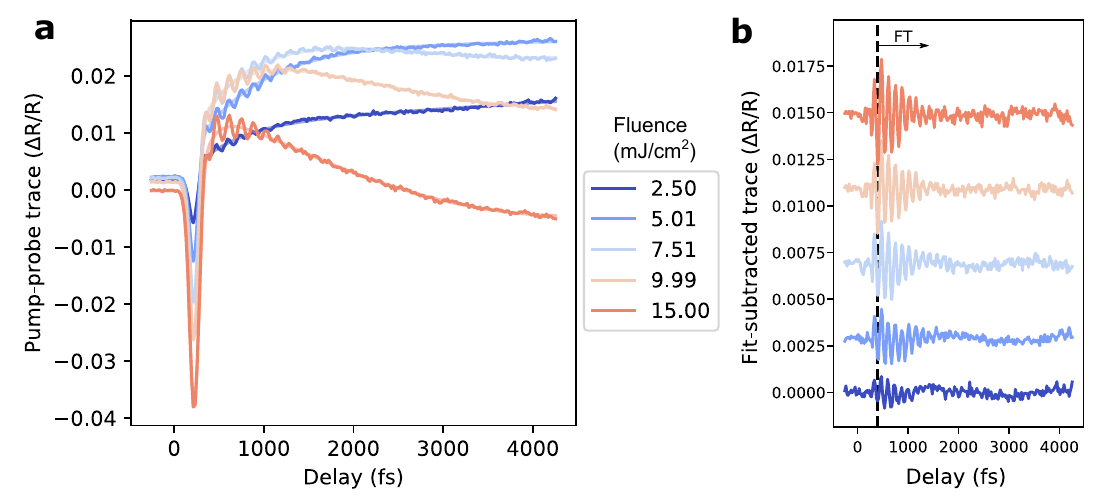}
\caption{(a) Full 1800\,nm probe data at 155\,K for the data shown in Fig.~4. Data taken in 700\,nm pump, 1800\,nm probe configuration. Also shown are the fits (faded lines below data), which are barely perceivable due to the quality of the fits. (b) Traces after background subtraction}
\label{SI_155K}
\end{figure*}

\end{document}